# Tai-e: A Static Analysis Framework for Java by Harnessing the Best Designs of Classics

Tian Tan
State Key Laboratory for Novel Software Technology
Nanjing University
tiantan@nju.edu.cn

Yue Li
State Key Laboratory for Novel Software Technology
Nanjing University
yueli@nju.edu.cn

**Abstract**

Static analysis is a mature field with applications to bug detection, security analysis, program understanding, optimization, and more. To facilitate these applications, static analysis frameworks play an essential role by providing a series of fundamental services such as intermediate representation (IR) abstraction, control flow graph construction, points-to/alias information computation, and so on. However, despite impressive progress of static analysis, and this field has seen several popular frameworks in the last decades, it is still not clear how a static analysis framework should be designed in a way that users (analysis developers) could benefit more: for example, what a good IR (for analysis) ought to look like? What functionalities should the module of fundamental analyses provide to ease or accelerate client analyses? How to develop and integrate new analysis conveniently? How to manage multiple analyses for a specific task?

To answer these questions, in this work, we discuss the design trade-offs for the crucial components of a static analysis framework, and argue for the most appropriate design by following the HBDC (Harnessing the Best Designs of Classics) principle: *for each crucial component of a static analysis framework, we compare the design choices made for it (possibly) by different classic frameworks such as* Soot, WALA, SpotBugs *and* Doop*, and choose arguably the best one, but if none is good enough, we then propose a better design*. These selected or newly proposed designs finally constitute Tai-e, a new static analysis framework for Java, which has been implemented from scratch. Specifically, Tai-e is novel in the designs of several aspects like IR, pointer analysis and development of new analyses, etc., leading to an easy-to-learn, easy-to-use and efficient system. To our knowledge, this is the first work that systematically explores the designs and implementations of various static analysis frameworks, and we believe it provides useful materials and viewpoints for building better static analysis infrastructures, and we expect it to draw more attentions of the community to this challenging but tangible topic.

***Keywords:*** static analysis, framework design, Java

## 1 Introduction

Static analysis is a well-studied technique that has been successfully applied to many applications like bug detection [12, 56], security analysis [3, 49], code optimization [65, 69], program understanding [46, 71] and verification [20, 58], and its effect has translated into real benefit for a substantial number of research work and industry products [15, 61]. To facilitate these applications (by implementing specific analysis algorithms), static analysis frameworks play an essential role by providing a series of fundamental services such as intermediate representation (IR) abstraction, control flow graph construction, points-to/alias information computation, and more. However, despite impressive progress of static analysis, and this field has seen several popular frameworks in the last decades [10, 26, 68, 77, 78], we are still not clear about what a good static analysis framework ought to look like, or at least, whether the designs of existing classic frameworks are good enough, and if not, can we have better ones?

This is a challenging problem, as system design is mostly a trade-off among different goals such as simplicity, efficiency and usability (one is often implemented at the expense of another); in addition, design intents are sometimes very subtle, which can hardly be aware of without the explanations from designers. But this does not mean that we cannot have some objective observations and reasonable judgments for this subjective problem. Despite very few, there is past work that attempt to offer some lessons learned for improving (or adding) certain facilities for their analysis frameworks after using it for a period of time [35, 63], but none addresses the core of our problem — we still lack a systematic view to examine the quality of a static analysis framework from the perspective of developers who rely on the framework to create new analysis; in other words, it is still unclear how a static analysis framework should be designed such that analysis developers can benefit more from it.

To deal with this problem, in this paper, we take one step forward by discussing the design trade-offs for each of the following crucial components that a static analysis framework (for Java) is supposed to provide.

- *Program Abstraction.* It needs to provide an abstraction model, including IR, type system, class hierarchy, etc., to represent all program elements that are ready for various static analyses to obtain;
- *Fundamental Analyses.* It ought to support fundamental facilities to allow analysis developers to operate on analysis-friendly structures, e.g., control flow and call graphs, enabling classic graph-based algorithms, and

- to utilize abstracted memory information, e.g., points-to/alias relations, to build sophisticated analyses;
- *New Analysis Development.* It is supposed to offer a mechanism to develop and integrate any new analysis, covering both fundamental ones like exception or reflection analyses, and clients like bug detectors or security analyzers;
- *Multiple Analyses Management.* It should provide a standardized approach to managing multiple analyses (e.g., configure the dependencies or cooperate the results of them) when they are required to work together for accomplishing a specific analysis task.

In addition, for each of the above crucial components, we argue for the most appropriate design by following the HBDC (Harnessing the Best Designs of Classics) principle:

*Given any component, we compare the design choices made for it (possibly) by different classic frameworks such as* Soot *[77],* WALA *[78],* SpotBugs *[68] and* Doop *[10], and choose arguably the best one. But if none is good enough, we then propose a better design.* These selected or newly proposed designs together constitute Tai-e, a new static analysis framework for Java, which has been implemented from scratch, and built with great care; these efforts finally contribute to *an easy-to-learn, easy-to-use* and *efficient* static analysis system.

Specifically, this work makes the following contributions.

**1.** We present the first work that systematically explores the designs and implementations of various classic static analysis frameworks for Java, and discuss their rationalities for different crucial analysis components, providing useful materials and viewpoints for building better static analysis infrastructures.

**2.** We introduce Tai-e, a new static analysis framework for Java, which is built from scratch, following the HBDC principle for every crucial component mentioned above. In addition to the *integration novelty* stemming from HBDC, Tai-e has its specific novel designs. For examples,

- Tai-e presents a usage-friendly IR for analysis implementation: compared to the IRs of Soot and WALA, it enables to produce more succinct code for implementing static analysis algorithms, and makes it easier to understand its underlying intents.
- Tai-e offers a more effective pointer analysis system by presenting some new designs (e.g., virtual-memory-like sparse bit sets to store points-to results) that can yield faster and more sound pointer analyses than all state of the arts for virtually all evaluated cases, in both context-insensitive and context-sensitive settings.
- Tai-e introduces a novel analysis plugin system to develop and integrate new analysis. A dozen analyses of Tai-e was developed on top of it, and it is currently being used by a number of ongoing internal projects; every developer agrees that this system can fulfill their practical needs and is simple to understand and apply.

Besides, despite not that novel, Tai-e is built with many engineering improvements. For example, in multiple-analyses management, Tai-e offers a scheme to more flexibly specify and resolve the dependencies among different analyses, and it enforces the analysis results to be stored distributively, enabling analysis developers to retrieve results consistently and conveniently. Such improvements can be found in every component of Tai-e. These engineering efforts, together with its designs (selected and newly proposed), making Tai-e easy-to-learn and easy-to-use.

**3.** We release Tai-e as an open-source framework for providing a uniform research platform to develop new fundamental analyses and clients *with low cost of learning and implementation* (https://github.com/pascal-lab/Tai-e). In addition, we develop an educational version of Tai-e (http://tai-e.pascal-lab.net/en/intro/overview.html) where eight assignments are carefully designed for systematically training students to implement various static analysis techniques to analyze real Java programs; this educational version was released three months ago (April 2022), and now it has attracted lecturers from 18 universities (for teaching purpose), team leaders from 10 companies (for training their engineers), and students from 85 universities and research institutes (for doing Tai-e assignments on our online judgment platform).

We will actively and constantly contribute to Tai-e by adding state-of-the-art analyses from existing work and developing new ones, including but not limited to incremental analysis [47, 48, 51], accumulation analysis [34], program slicing [24, 76], escape analysis [7, 14], new clients like bug detectors for data races [6, 48, 56], security analyzers for specific vulnerabilities [59, 63] (we now support taint analysis [21]), program understanding tools [46, 71], and more.

No doubt there is still a lot to explore, and years of research will be required to understand the right design choices and establish best practices for building an all-round and high-quality static analysis framework. Still, we believe that this work, together with Tai-e, have taken a big step on the way toward achieving this goal.

In the rest of the paper, we show how classic frameworks (if related) and Tai-e deal with the design points for each of the crucial components mentioned previously (Sections 2–5), discuss whether Tai-e, also as a system software, is reasonably designed in accordance with the classic STEADY principle [36, 37] from a view of system design (Section 6), discuss the related work and conclude the paper (Sections 7–8).

## 2 Program Abstraction

A static analysis framework needs to provide an abstraction model of programs, including IR, type system, class hierarchy, etc., to represent all program elements for static analysis to conveniently obtain. Below, we elaborate on some key design choices made by Tai-e and other related frameworks for this model and explain why we arrive at making them.

## 2.1 IR

We consider two classic frameworks, Soot and WALA, that have their specific IRs for analysis. Soot is virtually the most popular static analysis framework for Java over 20 years, and WALA also has attracted many researchers to build analyses during the last decade. The IR of Tai-e is largely inspired by these two frameworks with notable improvements for helping analysis developers implement more concise analysis code and better understand the underlying intents of IR. Below we take a representative example to illustrate the design choices made by these frameworks.

Soot represents all statements that have "=" operator inside as AssignStmt and does not explicitly distinguish the concrete cases of assign statements, e.g., new, load, store, unary, binary statements, etc. (but WALA and Tai-e do). Although utilizing fewer statement types makes IR simpler, it may also introduce many unnecessary conditional type checks. As an example, Figure 1 depicts how a binary statement (e.g., $x = y + z$), as a parameter of processBinary, is processed in different frameworks. The declared type of the assign parameter (line 2) has to be AssignStmt as it is the lowest-level interface in the class hierarchy of Soot to represent a binary statement; accordingly, conditional check to the right operand of assign is required (line 5) to ensure the casting (line 6) to be safe.

In addition, as Soot always returns Value (the highest-level interface in the class hierarchy) to represent data, e.g., local, constant, expression, reference, etc., further casting is needed to determine the right type of data. Line 8 explains one such case: the type of the left operand of a binary statement is a local variable (this is a pre-knowledge of Soot, but if users do not know it, a conditional type check is also required here), but as Soot declares Value as the returned type of getLeftOp(), casting to Local is needed due to type constraint. As another example (not shown in code), Soot provides <Value getCondition()> of class IfStmt to obtain the conditional expression of an if statement; that means Soot still returns Value even if it knows that the condition of an if statement must be a type of ConditionExpr (can be validated via its code), which is a subtype of Value, resulting in another case of unnecessary casting. The above cases imply that distinction in IR may appear subtle, but its effect is often profound.

WALA does not have this problem and it adopts a different strategy to represent local variables. For instance, getUse() in line 30 returns an int value (op2) as the index to access the information of the corresponding operand: to obtain the constant value stored in this operand, op2 is used as index to lookup a symbol table (line 39) which can be retrieved from ir (line 37); in other words, unlike Soot (line 10) and Tai-e (line 47) that directly holds the operand values, WALA adopts the above int index strategy to obtain the related values, which is not straightforward to understand and debug.

```
1  // Soot
2  void processBinary(AssignStmt assign) {
3      Value rightOp = assign.getRightOp();
4      // "Abstract" statement introduces conditional checks
5      if (rightOp instanceof BinopExpr) {
6          BinopExpr binopExpr = (BinopExpr) rightOp;
7          // "Top" returned type needs casting
8          Local lhs = (Local) assign.getLeftOp();
9          Value op1 = binopExpr.getOp1();
10         Value op2 = binopExpr.getOp2();
11         // obtain name of op1
12         if (op1 instanceof Local) {
13             Local var1 = (Local) op1;
14             String name1 = var1.getName();
15         }
16         // obtain type of op1
17         Type type1 = op1.getType();
18         // obtain constant value of op2
19         if (op2 instanceof Constant) {
20             Constant val2 = (Constant) op2;
21         }
22     }
23 }
24 // WALA
25 void processBinary(SSABinaryOpInstruction binary, IR ir) {
26     int lhs = binary.getDef();
27     // getUse() returns an int value as index to access the
28     // info of the corresponding operand: lines 32, 37-39
29     int op1 = binary.getUse(0);
30     int op2 = binary.getUse(1);
31     // obtain name of op1
32     String[] name1 = ir.getLocalNames(binary.iIndex(), op1);
33     // obtain type of op1
34     TypeInference tinfer = TypeInference.make(ir, true);
35     TypeAbstraction type1 = ti.getType(op1);
36     // obtain constant value of op2
37     SymbolTable symbolTable = ir.getSymbolTable();
38     if (symbolTable.isConstant(op2)) {
39         Object val2 = symbolTable.getConstantValue(op2);
40     }
41 }
42 // Tai-e
43 void processBinary(Binary binary) {
44     Var lhs = binary.getLValue();
45     BinaryExp binaryExp = binary.getRValue();
46     Var op1 = binaryExp.getOperand1();
47     Var op2 = binaryExp.getOperand2();
48     // obtain name of op1
49     String name1 = op1.getName();
50     // obtain type of op1
51     Type type1 = op1.getType();
52     // obtain constant value of op2
53     if (op2.isConst()) {
54         Literal val2 = op2.getConstValue();
55     }
56 }
```

**Figure 1.** An example for illustrating how to process binary statement in Soot, WALA and Tai-e based on different IRs.

Moreover, possibly due to the above design, different from `Soot` (lines 11–17) and `Tai-e` (lines 48–51) which uniformly obtains information from operands directly, `WALA` needs to utilize other interfaces to obtain the name and type of an operand, through `ir` (line 32) and `TypeInference` (lines 34–35) respectively, increasing learning costs.

`Tai-e` avoids the above issues of `Soot` and `WALA` as reflected in lines 43–56. Now we invite readers to view the code of Figure 1 in its entirety to experience how the same binary statement is handled differently based on different IRs. Given the binary statement example in Figure 1, and the fact that a program is often made of many different statements and expressions, we could anticipate considerable benefits afforded by `Tai-e`'s IR for writing more concise and understandable analysis code. Additionally, `Tai-e` introduces a few new IR designs to make it more accessible for certain analyses. For example, to facilitate pointer analysis, `Tai-e` associates each variable v with its related statements in its IR, and once v's value is changed during analysis, developers can directly and conveniently retrieve all the related statements of v via IR to take further actions.

We can hardly say that `Tai-e` offers a better IR than others as `Soot` and `WALA` may have different intents in their designs, which is difficult to be aware of without the explanations from designers. Still, we believe the IR of `Tai-e` reflects well its easy-to-learn/use nature and has clear advantages over others in that sense.

## 2.2 Type System and Class Hierarchy

A static analysis framework should provide a type system to represent types, and offer facilities for operating on types, e.g., for obtaining specific types and performing subtype checking. `Soot` explicitly distinguishes types like `primitive`, `reference`, `array`, `null` and `void` types, while `WALA` uses a uniform interface called `TypeReference` to represent all types. `Tai-e` is akin to `Soot` in designing type system as concrete type representation often eases API understanding and usage. For example, when we want to handle an `array` type which is passed as the parameter of a method, we can declare its (parameter) type as `ArrayType` (rather than `TypeReference`), then we know that this method can only accept `array` type and no further type checking is needed.

A static analysis framework should offer functionalities for managing classes, class members and their resolutions related to class hierarchy, e.g., obtaining a class by its name, returning subclasses of a given class, resolving fields based on field references and resolving methods according to method dispatch. Unlike `Soot` which places these functionalities in different classes like `Scene` and `FastHierarchy` (a bit distracting when learning the framework), `Tai-e` prefers `WALA`'s design which accesses all class related information through a single entry named `IClassHierarchy` (easier to find and maintain relevant APIs). Compared to `WALA`, `Tai-e` offers more facilities like getting class members by their signatures.

In a nutshell, `Tai-e` has no novelty for type system and class hierarchy, but it follows the HBDC principle to choose more appropriate designs and implement them uniformly.

## 3 Fundamental Analyses

Broadly, static analysis approximates how abstracted data flows along the control structure of a program according to the language semantics and runtime environment. Accordingly, a static analysis framework should offer fundamental facilities to produce such control structures like control flow graphs and call graphs (which are analysis-friendly structures that enable classic graph-based analysis algorithms) to develop various data flow analyses; besides, we need a pointer analysis to compute abstractions of the possible values/relations of pointer variables (points-to/alias information) in a program that are required by many other fundamental analyses and clients [65, 70]. In this section, we introduce what design choices are made by different frameworks for these two fundamental facilities: *pointer analysis* and *control/data flow analysis*, and explain how `Tai-e` is inspired by and differs from classic frameworks in designing them.

### 3.1 Pointer Analysis

"Pointer analysis is one of the most fundamental static program analyses, on which virtually all others are built." [40]. `Soot` offers a pioneer (context-insensitive) pointer analysis system for Java called `Spark` that is highly optimized and runs very fast [39], and `WALA` also implements a pointer analysis system with context-sensitivity enhancement [70]. `Doop` [10] is another classic pointer analysis framework that is full of clever and useful designs. Unlike `Soot`, `WALA` and `Tai-e` which are imperative (implemented in Java), `Doop` is fully declarative and implemented in Datalog, and it is considered as the mainstream platform to implement and compare different newly proposed pointer analysis algorithms for Java in the last decade [13, 27–30, 33, 41–43, 66, 67, 73–75]. All of these frameworks implement the same Andersen-style algorithm [1] as the core of pointer analysis; however, different choices are made by them for the following key points that need to be considered when designing a pointer analysis system for Java:

- a *representation* of points-to information
- a *solver* for propagating points-to information
- a *heap manager* for modeling heap objects
- a *context manager* for handling context sensitivity[1]

Besides, a practical pointer analysis system should offer a mechanism to conveniently develop new analyses (e.g., reflection analysis, exception analysis, taint analysis, etc.) that need to interact with pointer analysis. We next discuss how `Tai-e` and others deal with the above design points. For each point, only its major part is discussed due to limited space.

---

[1]Unlike pointer analysis for C/C++, context sensitivity is much more practically useful than flow sensitivity in improving precision for Java [65]

Tai-e: A Static Analysis Framework for Java by Harnessing the Best Designs of Classics

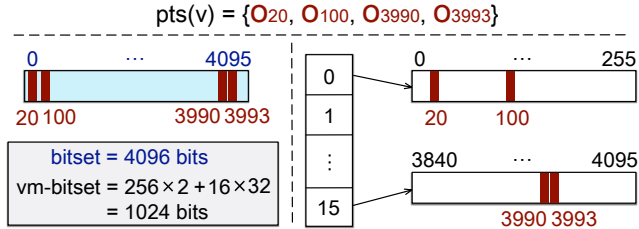

**Figure 2.** An example for illustrating regular ($2^{12}$ = 4096) and (one-level) "virtual memory"-like sparse bit sets in Tai-e.

***Representation of Points-to Information.*** Pointer analysis requires a uniform data structure to effectively represent points-to set that is associated with each variable in a program. Both Spark and WALA adopt a hybrid points-to set: when the size of set is less than certain value, they use array to store the pointed-to objects; otherwise, a regular bit set is considered to represent the points-to set. Tai-e follows this hybrid approach, but the bit set is designed differently.

Assume a regular bit set uses $2^{12}$ = 4096 bits to represent 4096 pointed-to objects stored in the points-to set of any variable that points to them. Figure 2 depicts a case that variable v has a points-to set $\{O_{20}, O_{100}, O_{3990}, O_{3993}\}$, and the four objects inside are presented by 1 (highlighted in red color) and the other bits remain 0 in the 4096-bits set. It is not hard to see that many bits are wasted when there are only a few pointed-to objects. To address this issue, Tai-e adopts a scheme called "virtual memory"-like sparse bit set [11] to represent points-to sets, as shown in Figure 2. In this case, 16 integers (each occupies 32-bits) are used as pointers for referring to maximally sixteen 256-bits sets (also 4096 bits in total). Since there are only four objects to represent in this case, two 256-bits sets are enough to store them, totaling 1024 = 2 × 256 + 16 × 32 (space for storing pointers) bits instead of the 4096 bits required by regular bit set.

In practice, to save more space, instead of the one-level page table used in the above example, Tai-e adopts two-level page table (like the concept in virtual memory) for referring to objects, and the table size is dynamically determined according to the number of pointed-to objects. Compared to regular bit set (used in Soot and WALA), the "virtual memory"-like sparse bit set in Tai-e helps save on average 23% (up to 40%) memory for context-sensitive pointer analysis in the experiment (introduced later). When we constrain the memory size to a limited one (like 8G for a laptop), this approach helps Tai-e scale for three more benchmarks under context-sensitive settings. Actually, the idea of sparse bit set is not new and it is used in some pointer analysis work for C/C++ but with different "sparse" strategies such as [4]. More attempts for designing new bit sets are encouraged for Java. Note that as Doop is declarative, its points-to set representation is not accessible to users, and different Datalog engines may use different representations [2, 10].

***Solver.*** Pointer analysis relies on its solver to propagate points-to information, and this process is usually based on a graph structure. For Spark, it is called pointer assignment graph (PAG) and its edges basically show how pointers are syntactically assigned to others. PAG is straightforward to understand (its nodes directly correspond to the code elements) and it is used in many analysis work [48, 51]. However, Tai-e adopts another structure called pointer flow graph (PFG) that is very similar to the flow graph in WALA; actually, the Datalog rules of Doop can also constitute such a graph although it does not encode it explicitly.

Unlike PAG, every node in PFG is a real pointer that points to objects. As shown in the example below, suppose variables a and b point to objects o1 and o2 (not shown) respectively. By the store operation a.f = b, o2 is propagated to o1.f, and the real pointer that points to o2 is o1.f, rather than the field reference a.f. Thus in PFG, we add the edge (from node b to node o1.f) that reflects the real points-to relation (means what are pointed to by pointer b will flow to pointer o1.f).

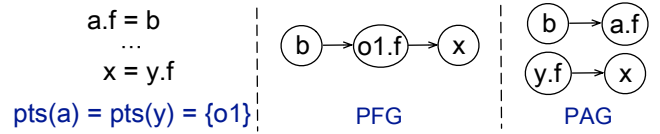

This design enables to directly retrieve the reachability information from PFG even in the presence of aliasing. In the above example, o2 will flow to x, if a and y are aliases (they point to the same object o1), and this reachable flow can be expressed via the path in PFG (but cannot in PAG). PFG can avoid the reiteration of the worklist algorithm that occurs in Spark (based on PAG) when handling field stores and loads, and it also eases to develop sophisticated pointer analysis algorithms that need direct object flow information [41, 43]

***Heap Manager.*** Typically, a huge number of heap objects will be allocated when executing a Java program, and as the protagonists of pointer analysis, how to model heap is an important issue that needs to be addressed. Regarding heap modeling, Tai-e is largely inspired by Doop and divide heap objects into four categories.

- *New object.* It includes all the objects that are allocated explicitly via keyword new (e.g., new T, new T[]).
- *Constant object.* It represents the constant objects whose fields will not be modified, (e.g., String, T.class, etc.). We can omit contexts for these objects in context sensitivity (no need to distinguish them via contexts as their field values remain the same through execution).
- *Merged object.* It covers the objects that could be merged (i.e., the new and constant objects with the same type) and Tai-e provides facilities to abstract them and access their origins, which eases the development of various heap abstraction techniques for pointer analysis [13, 19, 29, 75].

- *Mocked object.* It indicates the objects with no allocation sites in Java code. For instances, the metaobjects of reflection (e.g., Method/Field objects allocated in native code), the special objects created by JVM (e.g., main thread and parameters), taint values (used by taint analysis), and abstracted objects for modeling framework behaviors. Distinguishing mocked objects from others eases heap management, e.g., we can easily apply distinct heap contexts to mocked objects, and design specific tracing or logging strategies for them.

Both Spark and WALA consider the first two categories, and they additionally support merged objects but with limited capabilities: the merged objects are needed to be specified within a single method (WALA) or the entire program otherwise (Spark and WALA), and no other granularities are supported. As for Doop, Tai-e is mostly like it with some differences. For example, Tai-e regards all environment-related objects as mocked ones (including the objects of file system, main thread, system thread group, etc.) but Doop does not. These objects should be treated uniformly for consistency.

***Context Manager.*** Context sensitivity is the most widely used approach to improving precision of Java pointer analysis [65, 70], and we need a strategy to manage various context-sensitivity variants (call-site sensitivity [64], object sensitivity [54] and type sensitivity [66]) with different context lengths, for both method calls and heap objects.

Soot does not have an effective context-sensitive pointer analysis system: Spark is context-insensitive; Paddle is a BDD-based context-sensitive pointer analysis [38] of Soot and it has been shown to be noticeably less efficient than Doop [10] and hasn't been maintained for years. Doop provides a set of elegant rules to deal with context sensitivity. However, due to the limitation of Datalog, for each context length for the combination of method calls and heap objects, developers have to write a separate implementation for context-sensitive analysis, resulting in redundant code. In contrast, Tai-e is imperative and it can easily treat context length as an input parameter to the same implementation of context-sensitive analysis. WALA only provides context management for method calls, and its heap contexts directly inherit from the ones selected for the method that includes the allocation site of the heap object. Compared to WALA, Tai-e offers more flexible context management: developers can specify the contexts for both method calls and heap objects, e.g., 3-call-site (or 2-object) sensitivity for method calls and 1-call-site (or 1-object) sensitivity for heap objects.

Tai-e also provides facilities to develop selective context sensitivity (currently a hot research topic of Java pointer analysis) that scales for large and complex Java programs with good precision. Now many state-of-the-art selective pointer analyses like Zipper [41], Zipper$^e$ [43], Scaler [42] and Mahjong [75] have been implemented in Tai-e, serving as a uniform pointer analysis framework to compare and develop new context-sensitivity approaches.

***Experimental Evaluation.*** So far, we have explained how Tai-e makes choices for the key points for designing pointer analysis following the HBDC principle, to ease the development of sophisticated pointer analysis algorithms. Now we show how Tai-e performs in practice compared to the state-of-the-art pointer analysis frameworks.

*State of the arts.* We consider two state of the arts, Doop and Qilin [22] (the latest versions of both) which are representatives of declarative and imperative pointer analysis frameworks. Specifically, Qilin is a recently released tool developed on top of Soot, and some designs of it are similar to Spark (e.g., the same representation of points-to information and similar data structures used for solvers) but with enhancement of context sensitivity. As Spark does not support context sensitivity, it is not included in Table 1, but we will later summarize its results when explaining the data for context-insensitive pointer analysis. Unlike Tai-e, Doop, Qilin and Spark, WALA does not accept the output of dynamic reflection analysis as its input (explained in next paragraph), and it runs noticeably slower than others under context sensitivity. Therefore, it is excluded in our experiment.

*Benchmarks and Settings.* We consider standard Java DaCapo benchmarks [5] plus several large real-world applications that are often used in recent literature [29, 43, 50, 73]. We conduct all experiments on a machine with an Intel Xeon 2.2GHz CPU and 128GB of memory. All benchmarks are analyzed with a large Java library JDK 1.6 that is widely used in recent related work [22, 27–29, 31, 42, 43, 50]. All frameworks adopt the same reflection results for the same program by running the dynamic reflection analysis tool TamiFlex [8]. Time budget is set to three hours for each analysis.

*Evaluation metrics.* Pointer analysis algorithms are typically evaluated in the same framework to ensure the same soundness, so that efficiency and precision can be compared. To compare the effectiveness of pointer analyses provided in different frameworks, soundness and efficiency are the two most important metrics to evaluate whether program behaviors are well over-approximated and whether the analysis cost is reasonable, respectively. To evaluate soundness (or in some research fields, it is more understandable to say that whether an analysis is more complete than another), we conduct recall experiments to record the amounts of real reachable methods and call graph edges (The two "Total" columns in Table 1) that are dynamically collected when running the benchmarks.

*Results.* Table 1 shows the detailed results. Generally, Tai-e achieves higher recall (better soundness) than Qilin, Doop and Spark in both clients (amounts of reachable methods and call graph edges) for all programs. Specially, for these two clients, Tai-e's recall rates are 95.87% and 91.31% respectively on average, while Qilin's are 90.54% and 83.51%, Doop's are 78.10% and 68.44%, and Spark's are 81.32% and 73.20% (not shown in the table). We found that Doop is not taking full advantage of the reflection information offered by Tamiflex (e.g., omits classes mentioned in reflection logs from the initial facts upon which the analysis logic runs). Therefore, its recall results are not good for these cases.

**Table 1.** Pointer analysis results in terms of recall and analysis time. "Total", "Recall" and "R/T" mean the real results collected by running dynamic analysis, the real results resolved by a pointer analysis, and Recall/Total (recall rate), respectively. "#varpt", "#reach" and "#edges" mean the total number of points-to relations for all variables, reachable methods and call graph edges respectively. "2-obj" and "2-call" represent two widely adopted context-sensitive pointer analysis algorithms (2 levels of object sensitivity and call-site sensitivity). "–" means the analysis cannot finish running within time budget or run out of memory.

| Program | Tool | Reachable Methods | | | Call Graph Edges | | | Context Insensitivity | | | | 2-obj | 2-call |
|---|---|---|---|---|---|---|---|---|---|---|---|---|---|
| | | Recall | Total | R/T | Recall | Total | R/T | Time (s) | #varpt | #reach | #edges | Time (s) | Time (s) |
| findbugs | Tai-e | 5,808 | | 99.16% | 13,899 | | 98.75% | 15.2 | 7,355,719 | 17,353 | 107,339 | 1419.1 | 2114.5 |
| | Qilin | 5,663 | 5,857 | 96.69% | 13,323 | 14,075 | 94.66% | 23.5 | 7,360,275 | 16,988 | 106,998 | 2033.0 | 4214.6 |
| | Doop | 4,835 | | 82.55% | 11,121 | | 79.01% | 46.0 | 5,239,275 | 13,660 | 83,699 | 638.0 | 5292.0 |
| soot | Tai-e | 4,288 | | 98.08% | 16,233 | | 98.67% | 107.7 | 84,628,666 | 32,918 | 415,728 | – | – |
| | Qilin | 4,225 | 4,372 | 96.64% | 16,120 | 16,452 | 97.98% | 173.9 | 82,179,258 | 32,780 | 420,243 | – | – |
| | Doop | 4,253 | | 97.28% | 16,156 | | 98.20% | 454.0 | 74,564,948 | 32,600 | 413,411 | – | – |
| gruntspud | Tai-e | 14,360 | | 98.74% | 41,426 | | 98.40% | 69.0 | 48,179,683 | 39,800 | 274,872 | – | – |
| | Qilin | 14,279 | 14,543 | 98.18% | 41,261 | 42,099 | 98.01% | 113.9 | 50,073,466 | 39,247 | 273,794 | – | – |
| | Doop | 9,508 | | 65.38% | 25,363 | | 60.25% | 159.0 | 22,309,125 | 25,077 | 154,763 | – | – |
| columba | Tai-e | 6,673 | | 98.76% | 14,368 | | 97.81% | 135.8 | 107,856,623 | 56,787 | 425,149 | – | – |
| | Qilin | 6,617 | 6,757 | 97.93% | 14,260 | 14,689 | 97.08% | 170.0 | 101,298,328 | 56,726 | 416,791 | – | – |
| | Doop | 4,983 | | 73.75% | 9,798 | | 66.70% | 616.0 | 70,315,191 | 42,665 | 286,990 | – | – |
| antlr | Tai-e | 2,697 | | 96.81% | 10,086 | | 97.53% | 9.0 | 2,003,409 | 8,674 | 59,190 | 37.1 | 654.0 |
| | Qilin | 2,568 | 2,786 | 92.18% | 9,531 | 10,341 | 92.17% | 11.5 | 2,066,405 | 8,363 | 58,486 | 71.0 | 2009.9 |
| | Doop | 2,510 | | 90.09% | 9,401 | | 90.91% | 31.0 | 2,385,993 | 8,250 | 57,560 | 282.0 | 1646.0 |
| bloat | Tai-e | 3,339 | | 97.60% | 11,794 | | 98.05% | 11.7 | 3,270,010 | 9,936 | 69,219 | 452.3 | 1107.0 |
| | Qilin | 3,292 | 3,421 | 96.23% | 11,722 | 12,029 | 97.45% | 13.1 | 3,363,502 | 9,631 | 68,922 | 1253.8 | 2691.2 |
| | Doop | 3,231 | | 94.45% | 11,297 | | 93.91% | 26.0 | 3,393,681 | 9,509 | 67,604 | 1394.0 | – |
| xalan | Tai-e | 3,826 | | 96.47% | 9,254 | | 96.26% | 9.8 | 2,805,148 | 12,942 | 72,661 | 1854.0 | 1221.5 |
| | Qilin | 3,650 | 3,966 | 92.03% | 8,571 | 9,614 | 89.15% | 16.5 | 2,697,455 | 12,526 | 70,612 | 869.3 | 2546.6 |
| | Doop | 3,132 | | 78.97% | 6,999 | | 72.80% | 28.0 | 1,681,360 | 9,956 | 53,278 | 439.0 | 1359.4 |
| eclipse | Tai-e | 7,080 | | 87.48% | 18,050 | | 86.30% | 26.8 | 22,633,879 | 23,920 | 184,294 | – | – |
| | Qilin | 6,936 | 8,093 | 85.70% | 17,476 | 20,916 | 83.55% | 48.1 | 24,230,566 | 23,550 | 184,290 | – | – |
| | Doop | 3,378 | | 41.74% | 6,860 | | 32.80% | 21.0 | 2,044,456 | 9,764 | 56,595 | – | – |
| hsqldb | Tai-e | 2,608 | | 95.43% | 5,856 | | 93.03% | 11.7 | 1,838,669 | 11,588 | 64,393 | – | 654.6 |
| | Qilin | 1,707 | 2,733 | 62.46% | 2,996 | 6,295 | 47.59% | 8.8 | 1,016,624 | 7,570 | 42,896 | – | 968.3 |
| | Doop | 1,626 | | 59.50% | 2,841 | | 45.13% | 78.0 | 839,150 | 6,945 | 37,414 | – | 2477.4 |
| jython | Tai-e | 4,243 | | 87.76% | 11,639 | | 32.36% | 17.0 | 10,885,453 | 13,076 | 121,603 | – | – |
| | Qilin | 4,056 | 4,835 | 83.89% | 10,847 | 35,970 | 30.16% | 25.7 | 10,101,462 | 12,650 | 119,872 | – | – |
| | Doop | 3,985 | | 82.42% | 10,717 | | 29.79% | 78.0 | 11,267,651 | 12,507 | 118,706 | – | – |
| luindex | Tai-e | 2,179 | | 96.16% | 4,461 | | 94.71% | 8.8 | 829,314 | 7,899 | 41,520 | 18.4 | 515.8 |
| | Qilin | 2,052 | 2,266 | 90.56% | 3,907 | 4,710 | 82.95% | 9.1 | 889,969 | 7,590 | 40,828 | 30.4 | 1226.4 |
| | Doop | 1,973 | | 87.07% | 3,376 | | 71.68% | 14.0 | 921,998 | 7,456 | 39,743 | 46.0 | 469.2 |
| lusearch | Tai-e | 1,736 | | 95.07% | 3,261 | | 92.88% | 9.2 | 981,532 | 8,574 | 44,839 | 18.9 | 509.8 |
| | Qilin | 1,606 | 1,826 | 87.95% | 2,705 | 3,511 | 77.04% | 9.5 | 1,048,650 | 8,263 | 44,166 | 47.7 | 1234.2 |
| | Doop | 1,545 | | 84.61% | 2,534 | | 72.17% | 15.0 | 1,099,851 | 8,136 | 43,072 | 49.0 | 1435.8 |
| pmd | Tai-e | 3,908 | | 97.09% | 9,287 | | 96.39% | 10.1 | 2,848,296 | 13,311 | 73,890 | 39.1 | 1174.2 |
| | Qilin | 3,735 | 4,025 | 92.80% | 8,613 | 9,635 | 89.39% | 15.8 | 2,974,963 | 12,981 | 73,292 | 75.0 | 2136.0 |
| | Doop | 2,718 | | 67.53% | 5,737 | | 59.54% | 29.0 | 1,330,917 | 8,905 | 46,862 | 68.0 | 1605.9 |
| chart | Tai-e | 5,197 | | 97.58% | 12,751 | | 97.26% | 17.4 | 5,755,512 | 16,455 | 87,938 | 184.9 | 856.7 |
| | Qilin | 5,024 | 5,326 | 94.33% | 12,064 | 13,110 | 92.02% | 21.8 | 6,103,008 | 16,109 | 87,848 | 283.6 | 2281.8 |
| | Doop | 4,691 | | 88.08% | 11,172 | | 85.22% | 42.0 | 4,679,309 | 13,643 | 73,106 | 216.0 | 3506.0 |

Many reasons could possibly contribute to the good recall of Tai-e, and it is very hard to accurately find them out. One explanation is that `Tai-e` may have better static treatment to language features such as reflection resolution (we have found such cases), native code modeling, etc., and perhaps the implementation of analysis algorithms in `Tai-e` is more robust and thus more dynamic behaviors are well over-approximated.

Meanwhile, `Tai-e` has the fastest analysis speed compared to other frameworks for virtually all cases under both context-insensitive and context-sensitive settings, as reflected in the three "Time(s)" columns in Table 1. For context sensitivity, we consider two widely used algorithms: object sensitivity and call-site sensitivity, both of which have context lengths of two [43, 65, 67, 74, 75], denoted as 2-obj and 2-call, respectively. Note that for 2-call, the latest version of `Doop` either runs out of memory (and killed by operating system) or exceeds time limit for all programs, we instead list the results of running an old version of `Doop` for 2-call in Table 1.

Now let us examine the analysis time. Note that we cannot easily draw a conclusion that one framework has better efficiency ability than others if it spends less time, as the soundnesses achieved by different frameworks vary (an analysis is faster perhaps because it analyzes less code than others). But we can trust the capability of `Tai-e` for yielding highly efficient pointer analysis (benefiting from its designs for efficiency like representation of points-to information and solver), because it achieves better recall (more sound) for all programs than all other frameworks; moreover, in many cases, even if `Tai-e` analyzes more reachable methods (#reach) and computes more points-to relations (#varpt) or call graph edges (#edges), it still runs faster than others (this also includes the case of `Spark`: `Tai-e` outperforms `Spark` in analysis speed for 12/14 programs even if `Tai-e` has significantly better recall than `Spark`), demonstrating again the good efficiency of `Tai-e`.

### 3.2 Control/Data Flow Analysis

The algorithms for building control and data flow analyses (e.g., control flow graph construction and various data flow analyses like constant propagation, live variables analysis, etc.) are standard and well understood in the field of compilers. However, as the providers that support fundamental facilities to build these analyses, static analysis frameworks may adopt different strategies in design details. We next introduce how `Tai-e`, `Soot`, `WALA` and `SpotBugs` make choices for certain key design points to ease the development of effective control and data flow analyses (`SpotBugs` is a famous static bug detection framework (the successor of `FindBugs` [25]) that has a powerful system of control/data flow analysis [68]).

***Control Flow Analysis.*** Building control flow graph (CFG) is the major task of control flow analysis, and despite its basic algorithm is standard, its effectiveness for facilitating users to build analysis varies. We use two examples to explain.

*Edge categories.* Unlike `Tai-e` and `SpotBugs`, `Soot` and `WALA` do not category edges of CFG, such as `IF_TRUE`, `IF_FALSE` and `CAUGHT_EXCEPTION`, etc. These well-categorized edge information will ease to develop certain analyses like path-sensitive and branch-correlated analysis, or perform exception-specific handling. Compared to `SpotBugs`, `Tai-e` provides additionally useful edge information, e.g., it labels the `switch` edges with case values, and the `exception` edges with concrete exception types. Note that developers can also parse out the edge information in `Soot` and `WALA` by resolving the related nodes and IR in their analyses, but doing so would be inconvenient and not easy to use.

*Exception handling.* Regarding CFG for Java, an essential factor that affects its effectiveness is how to statically resolve Java exception, and it is considered as the most critical source of unsoundness in program analysis that developers indicated should not be overlooked [15]. There are explicit and implicit exceptions: the former is thrown by `throw` statement and is caught by the corresponding `catch` statement if their types are matched, and the latter is thrown implicitly by JVM (like out of memory errors, and arithmetic exceptions). A sound CFG should consider both of them, but in many cases, there is possibly a huge number of implicit exceptional control flows in a program and they do not interact much with the normal flows (explicit exceptional control flows do as logics are often implemented in their `catch` body), and thus consider them in CFG may reduce analysis precision (and usability). Unlike `WALA` and `SpotBugs`, `Tai-e` and `Soot` distinguish explicit and implicit exceptional control flows and allow users to decide which ones should be added to a CFG. Moreover, `Tai-e` implements state-of-the-art exception analysis [9, 32] that resolves exception significantly more precise (sometimes also more sound) than `Soot`, `WALA` and `SpotBugs`, and offers it as an option for constructing CFG.

***Data Flow Analysis.*** Typically, to implement customized analysis, developers should follow the interfaces provided by a data flow analysis system to specify (1) data facts abstraction and initialization, (2) the transfer function for approximating different statements, and (3) the meet/join operation for merging data facts at control flow confluences, while making their analyses monotonic and safe-approximated. Below, we take three examples to illustrate how different designs may lead to different perceptions of use for developers.

*Data facts initialization.* Unlike `Tai-e`, `SpotBugs` and `Soot`, `WALA` does not allow to initialize data facts in the analysis; instead, it puts the related API in the solver, and thus every time we write a new analysis that needs different initialization, we have to additionally implement a new solver to override the API that is responsible for initializing data facts. We argue that an elegant design is to have just one solver to drive multiple data flow analyses, so that developers only need to focus on the implementations of their analyses (no need to know the details of solver).

Tai-e: A Static Analysis Framework for Java by Harnessing the Best Designs of Classics

*Edge transfer functions.* Unlike Tai-e, SpotBugs and WALA, Soot does not explicitly support edge transfer functions. Edge transfer functions differ from node transfer functions in that they allow distinct data facts to be sent to various successors of a particular node along the edges between these nodes, utilizing branch information (e.g., fact D is propagated when expression E's value is true) to create more effective analysis. When the body of edge transfer function is left empty, it is regarded as an identity function to directly propagate the OUT fact of its source node to the IN fact of its target node (i.e., at this point, the analysis will change back to the normal one where only node transfer function is in charge). However, in Soot, to leverage branch information, developers need to extend a special analysis called BranchedFlowAnalysis and implement the logics for both edges and nodes in its node transfer function, which is inconvenient and a bit cumbersome in design.

*Data facts comparison.* Unlike Tai-e and WALA, in SpotBugs and Soot, in order to determine whether data facts have changed during consecutive iterations and accordingly decide whether to add the relevant nodes to worklist for further handling, their analysis solvers retain the old data facts and will compare them to the new ones in each iteration in the same way. However, although optimization can be used for certain analyses, it is hard to accomplish when the solver is in charge of the comparison. In contrast, Tai-e and WALA reverse the comparison right to developers (actually, to the transfer function) and the solver waits for transfer function's decision before proceeding. For example, assume that we have 1000 variable-value pairs in the data fact of constant propagation. After solver sends the merged IN and old OUT facts (of certain node) to transfer function, constant propagation can directly update the old OUT fact (as new OUT fact) when it finds, e.g., a variable's value has changed from 1 to NAC, and then notifies the solver to add the successors of the node to worklist. As a result, the solver does not need to retain a copy of old fact and compare it (1000 pairs) to the new one.

## 4 New Analysis Development

A static analysis framework should offer mechanisms to incorporate new analyses, from intraprocedural to interprocedural ones, and it is important to keep in mind that working with pointer analysis is necessary for the creation of a variety of analyses [65, 70]. We introduce how Tai-e and other frameworks design for developing new analyses that need to interact with pointer analysis (Section 4.1), as well as for developing or integrating other analyses (Section 4.2).

### 4.1 Develop Analysis that Interacts with Pointer Analysis

Numerous static analyses, including fundamental ones like reflection analysis [44, 45] and exception analysis [9, 32] as well as clients like bug finders [12, 56] and security analyzers [3, 21, 49], require to interact with pointer analysis.

Doop naturally supports such interactive analysis and is able to yield elegant implementation, benefiting from Datalog's declarative ability. However, Doop is also limited by Datalog in implementing analysis that requires non-set-based lattices [52, 72], and it is hard to optimize specific analysis as Datalog solver adopts analysis-independent data structures and execution strategy [23]. As a result, imperative frameworks that facilitate such interactive analysis are in high demand. As representatives, Soot lacks this backing, whereas WALA does.

WALA provides a scheme to add new analysis that interacts with pointer analysis, but in a limited way. Briefly, developers need to implement an interface called ContextSelector to specify related call sites (of certain APIs) which the new analysis aims to model based on the points-to results; for example, to analyze reflective call v = c.newInstance(), developers encode ContextSelector to identify this call site and retrieve the Class objects, say C0, that are pointed to by c via pointer analysis. Then, developers need to implement ContextInterpreter to generate different fictitious but effect-equivalent IRs (e.g., v = new T(); v.<init>(): allocating an object and calling its constructor in this case) according to the resolved types of C0 (say T). Then these generated IRs are fed back to pointer analysis to continue the resolution for this reflective call.

We argue that this scheme has some limitations to facilitate interactions with pointer analysis. First, for certain analyses, it is insufficient to interact with pointer analysis by only giving developers a way to concentrate on and resolve related call sites; capabilities for monitoring the points-to information of specified variables are required. For example, in exception analysis, if the points-to set of variable e in throw e is changed, the points-to set of the corresponding catch variable should also be updated. Second, in many situations, it is simpler to update points-to-information or call graph edges directly, which can prevent the generation of excessive amounts of fictitious IR code and the need to invoke the solver to reanalyze the created code. Thus, a framework is supposed to also offer a mechanism to perceive changes for any variables of a program in order to accommodate more analyses; additionally, it should provide a way to directly adjust the points-to results and call graph edges for an easier or more effective engagement with pointer analysis.

Recently, Helm et al. [23] present an ambitious approach to collaborating various analyses on the fly, even with different analysis lattices. However, this approach is too complex to adopt for addressing our problem, because many constraints must be put in place by developers, including encoding the rules that govern how control engine should interpret and communicate the results of one analysis to others.

To practically facilitate the development of new analysis that needs to interact with pointer analysis, in Tai-e, we introduce a simple yet effective method called *analysis plugin system*. Currently a dozen analyses in `Tai-e` are built on top of this system, including fundamentals like reflection and exception analyses, clients like taint analysis, utility tools like analysis timer and constraint checker (for debugging), modern language feature handling like lambda expression and method reference analyses, runtime environment modeling like native code and thread modeling, and so on.

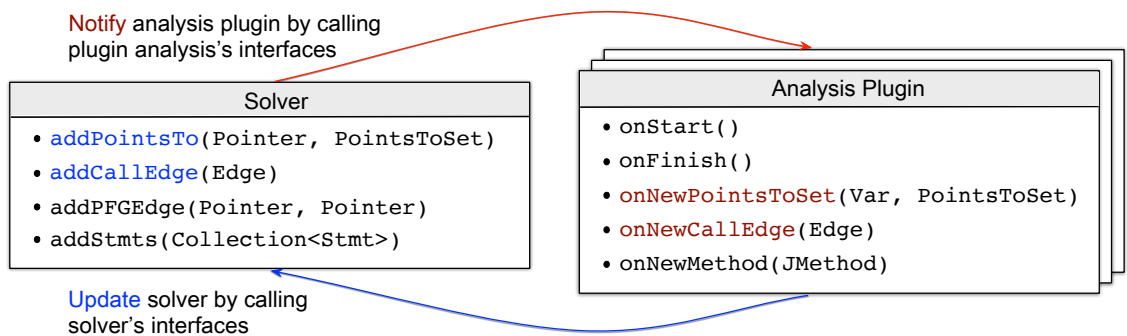

**Figure 3.** Overview of the analysis plugin system of `Tai-e` for developing new analyses, e.g., taint analysis, exception analysis, reflection analysis, etc. that require interaction with pointer analysis.

***Basic Idea.*** We explain how this analysis plugin system works. As shown in Figure 3, this system includes a pointer analysis solver and a number of analyses that communicate with it. Each of these analyses is referred to as an analysis plugin that needs to implement interface `Plugin` of `Tai-e`. The interactions between pointer analysis solver and analysis plugin are carried out by calling each other's APIs. The *core* APIs of `Solver` and `Plugin` are highlighted in blue and red, respectively. The `Solver` APIs have been implemented in the framework, and developers only need to implement the related APIs of `Plugin` to develop new analysis. The additional auxiliary APIs are optional and designed to make it easier to create specific functionalities; for example, `addStmts` of `Solver` can be called to simulate the effect of specific call sites, which is similar to the generated-IR approach of `WALA` mentioned above.

Let us briefly illustrate the basic working mechanism that drives those core APIs. Assuming you are implementing the `onNewPointsToSet` method of an analysis `Plugin`, this means whenever an interested variable's (parameter `Var`) points-to set (parameter `PointsToSet`) is changed (i.e., it points to more objects), you need to encode your logic to reflect the side effect made by this change; the final consequence of such an effect, from the perspective of pointer analysis, is to modify the points-to set of any related pointers or to add call graph edges at pertinent call sites. Accordingly, you should call the `addPointsTo` or `addCallEdge` methods of the `Solver` to alert it of these modifications. Conversely, during each analysis iteration, the `Solver` calls the `onNewPointsToSet` and `onNewCallEdge` methods of every

`Plugin` to notify them of any changes to the variables' points-to sets or call graph edges, respectively. As a result, to add a new analysis that interacts with pointer analysis, developers just need to implement a few methods of `Plugin` in accordance with the requirement, as previously described.

***Case Study.*** To better understand how to build new analyses on top of this plugin system, let us look at a specific example: creating a taint analysis, and its pseudocode is depicted in Figure 4. We recommend following the figure together with our text explaining the scenario below.

```
1 String s1 = x.source();
2 s3 = s2.concat(s1);
3 y.sink(s3);
```

The code above outlines a typical case for taint analysis: Sensitive data from an object, say `o1`, returned by `x.source()`, is combined with an object, say `o3`, pointed to by `s3`, returned by `s2.concat(s1)`. Finally, taint analysis reports that line 3 contains a leak of `o1`'s sensitive data because `o3` is used as an argument by a sink method.

As taint analysis is developed as an analysis plugin, it needs to implement some of its methods in Figure 3. In our case, as shown in Figure 4, two core methods `onNewCallEdge` (line 5) and `onNewPointsToSet` (line 16), and an auxiliary method `onFinish` (line 27) are considered.

Let us first examine `onNewCallEdge`. Assuming we are handling call site 1: `String s1 = x.source()` in the code snippet above, and for this call site, the pointer analysis solver has resolved a new call graph edge from edge source (which is a call site denoted as `edge.cs`), i.e., call site 1, to edge target, i.e., the dispatched method `source`; then the solver notifies the taint analysis plugin and passes this new edge to parameter `edge` in line 5, by calling plugin's `onNewCallEdge` method (the solver code is omitted).

Then the plugin code checks what the `edge.target` is.

If the target is a sensitive source method, denoted as `source` (line 6) (the `source`, `transfer` and `sink` methods are specified in the configuration file of taint analysis), a Tai-e: A Static Analysis Framework for Java by Harnessing the Best Designs of Classicsmocked taint object is created (line 7) and the plugin updates the solver that the left-hand side variable of this call site, namely `s1` (of the above code), should point to the taint object that was just created (line 8).

If the target is a transfer method (line 10), e.g., the `concat` method at call site 2: `s3 = s2.concat(s1)`, by which taint objects could be transferred from its parameters (e.g., `s1`) to other variables at the call site (e.g., `s3`) [21], for each of such transfer relation, denoted as `(from, to)` in line 11, the taint objects that are pointed to by `from` should be added in the points-to set of `to` (line 12 and lines 21 – 24). As a result, `s3` now points to the taint object transferred from `s1`. Actually, this transfer relation is also recorded in a data structure called `transferVars` (line 13), which is equivalent to add a taint-relevant flows-to edge summarized for transfer method. Then the code of `onNewPointsToSet` (line 16) is easy to understand: whenever a new taint object flows to parameter `v`, and `v` is also the `from` variable recorded in `transferVars` mentioned above, the corresponding flows-to variable `to`, should also point to this taint object (line 18) for soundness.

`onFinish` (line 27) will be called by solver after it reaches the fixed point. For each sink method and its sensitive parameter `param_i` (line 28), `onFinish` identifies all its call sites `cs` and checks to see if any taint objects flow to `param_i` (lines 29 – 31), and if they do, it reports this taint information.

Finally, let us discuss a possible design trade-off for this analysis plugin system. As described above, `Solver` invokes methods `onNewPointsToSet` and `onNewCallEdge` of every `Plugin` in turn to notify them any changes to variables' points-to sets or call graph edges. One might wonder why not allow each analysis plugin to register their interested variables or call graph edges to the solver in advance, so that the plugins only receive notifications when the registered variables or edges change. We do not take this design into consideration for three reasons. First of all, this will add more API to the interface to register the interesting variables, increasing developers' workload. Second, while creating an analysis plugin, it can be hard to identify and specify the call graph edges that are of relevance in advance. Third, we use runtime profiling and discover that our current design (the solver notifies every plugin in turn) actually introduces negligible cost in practice. As Section 6 will discuss, our analysis plugin system has received very positive user feedback.

### 4.2 Develop Other Analyses

A static analysis framework would be beneficial if it could also provide facilities for developing other analyses, including intraprocedrual and interprocedural ones.

Soot categories analysis into two kinds, method-level and program-level, and developers need to implement so-called `BodyTransformer` and `SceneTransformer` to develop their

```
1  class TaintAnalysis implements Plugin {
2  
3    Set<Pair<Var, Var>> transferVars;
4  
5    void onNewCallEdge(Edge edge) {
6      if edge.target is source:
7        o = heapModel.getMockObj("TaintObj", edge.cs)
8        solver.addPointsTo(edge.cs.lhsV, o)
9  
10     if edge.target is transfer:
11       foreach (from, to) in getTransfer(edge.target,
             edge.cs):
12         transferTaint(solver.getPointsToSet(from), to)
13         record (from, to) in transferVars
14   }
15  
16   void onNewPoinstToSet(Var v, PointsToSet pts) {
17     foreach (v, to) in transferVars:
18       transferTaint(pts, to)
19   }
20  
21   void transferTaint(PointsToSet pts, Var to) {
22     foreach o in pts:
23       if o.desc is "TaintObj":
24         solver.addPointsTo(to, o)
25   }
26  
27   void onFinish() {
28     foreach (sink, param_i) in sinks:
29       foreach callsite cs in solver.getCallersOf(sink):
30         foreach o in solver.getPointsToSet(cs.args[param_i]):
31           if o.desc is "TaintObj":
32             report(o, cs)
33   }
34 }
```

**Figure 4.** Plugin of taint analysis (pseudocode version).

specified intraprocedrual and whole-program analyses, respectively. SpotBugs facilitates to develop method-level and class-level analyses, and the latter covers certain field- and hierarchy-related analyses that require the information of classes. For example, `FindMaskedFields` is to detect if a field has the same name with local variables in its declaring class or its parents' fields; `RedundantInterfaces` is to find out if an interface implemented by a class has also been implemented by its parent. WALA does not have an explicit scheme to develop and integrate new analysis as it is often used as a library, and when users develop new analysis, they often need to specify their `main()` behaviors nearly from scratch.

Tai-e supports all three types of the analyses mentioned above (i.e., method-, class- and program-level analyses); furthermore, unlike Soot where developers require to follow the implicit rules to hard-code new analysis into frameworks, Tai-e (similar to SpotBugs) allows to configure new analysis in a decoupled way. To develop a new analysis in Tai-e, developers only need to choose one of the classes `MethodAnalysis`, `ClassAnalysis` and `ProgramAnalysis` to extend, and implement the corresponding `analyze()` method based on the specific application, and then register the analysis in a configuration file (specifying analysis name, dependent analyses, and certain default options if needed). Tai-e will take care of the process regarding how to automatically drive new analyses and record analysis results according to the specified analysis categories, dependences and options, which greatly reduces the effort required from developers.

## 5 Multiple Analyses Management

In many circumstances, an analysis depends on the outcomes of other analyses, and thus it will be helpful if the framework can provide a mechanism to coordinate multiple analyses. In this section, we discuss two crucial issues related to managing numerous analyses: how to configure an analysis and its dependencies (Section 5.1), and how to save the outcomes of one analysis and access them in another (Section 5.2)?

### 5.1 Configure Analysis and Its Dependences

WALA has no explicit management for multiple analyses, and as mentioned previously, Soot needs to hard-code new analyses while Tai-e and SpotBugs support to register them in the framework through configuration files and then drive them via reflection. Figure 5 depicts a configuration example for specifying control flow graph (CFG) builder in Tai-e with format of YAML [80] (a concise and human-friendly format that is commonly used for configuration files where data is being stored or transmitted).

```
- description: intraprocedural control-flow graph
  analysisClass: pascal.taie.analysis.graph.cfg.CFGBuilder
  id: cfg
  requires: [ throw(exception=explicit|all) ]
  options: # default values
    exception: explicit # | null | all
    dump: false # dump .dot files
```

**Figure 5.** A configuration example of Tai-e for CFG builder.

The first three attributes are easy to understand and the fourth one, requires specifies all the dependent analyses for cfg (i.e., all the analyses whose results are required by cfg). In this instance, it is an exception analysis (denoted by throw) and the boolean expression inside its followed parenthesis indicates that the dependent analysis is only required when the expression value is true. In our case, that means Tai-e needs to run throw before cfg only when option exception's value is explicit or all. The last attribute options lists all the default values for the options designed for cfg: only the explicit edges of exception are considered, and the analysis results are not dumped to files.

In Soot, before running an analysis, users must explicitly list any dependent analyses (including those that depend on the dependent ones) in the command. This approach is cumbersome and prone to mistakes for users who are unfamiliar with the framework.

In Tai-e and SpotBugs, the dependency resolution is automatic by analyzing configuration files, ensuring the correctness of execution order for all dependent analyses; besides, this approach frees up developers to concentrate on the specification of their own analysis, and saves their effort of writing command options when running an analysis.

Compared to SpotBugs, Tai-e is more flexible in resolving analysis dependences. Let us take an example to explain. We want to build two different types of CFGs: one that resolves a program's explicit exceptional edges and the other that ignores exception. In Tai-e, this can be done by simply giving different commands: one is "-a cfg" (-a means analysis) and the other is "-a cfg=exception:null". The latter uses runtime option value (exception:null) to override the default one (specified by exception:explicit in Figure 5). But in SpotBugs, as it does not support conditional logics to describe analysis options and dependences, every time we perform the analysis, we have to find the relevant configuration files/options and modify the values as necessary.

In summary, in order to facilitate simple usage, maintenance, and troubleshooting in terms of configuring and conducting analyses, Tai-e strives to guide users to modify code or configurations as little as possible.

### 5.2 Store/Access Analysis Results

Storing and accessing analysis results may seem like a minor concern that doesn't need to be discussed, but a good design, despite the fact that it may not look technical, can nevertheless produce a favorable user experience.

Unlike Tai-e and SpotBugs, WALA and Soot do not have a uniform mechanism to manage analysis results for all analyses (Soot only stores the results of some analyses like pointer analysis in the singleton instance of Scene). In SpotBugs, users need to remember different methods to get related results. For example, assuming you are implementing a dead code analysis for any given method (or IR, as IR is typically related to each method), to obtain the analysis result yielded by an intraprocedural analysis like CFG builder for this method, users should call IAnalysisCache.getMethodAnalysis(), and send it two arguments, one for the analysis they want (CFG builder) and the other for the method they wish to analyze. Accordingly, users need to remember and call another method and provide additional relevant arguments if they require the results from a class-related analysis.

However, in Tai-e, regarding analysis results, users only need to remember one method getResult(id) (id is the analysis name) for all types of analyses, including method-, class- and program-level analyses, as described in Section 4.2. For the same example above, as it is an intraprocedural or method-level analysis (which means that in Tai-e, it will extend MethodAnalysis and encode its analyze(ir) method where ir represents the IR for any method, and is passed by the framework), users only need to use the parameter ir to directly obtain the result by calling ir.getResult("cfg").

Similarly, if users are implementing a class-level analysis (which means it extends class `ClassAnalysis` and implements method `analyze(jclass)`), they only need to call `jclass.getResult(id)` to get the results. The straightforward user interface for accessing analysis results benefits from the fact that `Tai-e` automatically stores analysis results in different locations based on various types of analyses. As a result, users no longer have to bother trying to memorize complicated methods and specify additional arguments.

## 6 A View of System Design

Our primary goal is to make `Tai-e` an easy-to-learn and easy-to-use static analysis framework for Java, and to do this, we painstakingly design and implement it from scratch while adhering to the HBDC principle. From another perspective, `Tai-e` is a sophisticated system software, and a good system should also go by a few well-established design principles, which are summed up as STEADY by Butler W. Lampson in his seminal work for system design [36, 37]. In this section, we examine `Tai-e` from the view of STEADY which stands for "Simple, Timely, Efficient, Adaptive, Dependable and Yummy" [2].

***Simple.*** Keep the design simple. `Tai-e` follows the simplicity principle from the high-level goal to the low-level implementation. "Do one thing well" and "be aware of universal goals' [37]: `Tai-e` is designed to analyze Java programs only (there is no requirement for code transformation and for supporting to analyze other languages), so the designs for program abstractions and facilities to build analysis can be simplified. The code structure of `Tai-e`, including its modules or packages, is well organized and simple to comprehend. The facilities offered by `Tai-e`, accessed through its APIs, are likewise designed in an easy-to-use way. For example, as explained previously, the code for utilizing IR and retrieving type/class information (Section 2), developing certain sophisticated control/data flow analyses (Section 3), obtaining results of various analyses (Section 5), is simpler to write than others. In the lowest level, we even make an effort to make the code simple to understand. All of these factors, along with the emphasis placed on the adaptivity principle (discussed later), make `Tai-e` easy-to-learn and easy-to-use.

***Efficient.*** Build a high-performance system. As explained previously, as pointer analysis is a fundamental analysis, on which virtually all others are built [40], it severs as the framework's core motor, propelling other analyses, and thus its efficiency is of utmost importance. As demonstrated in a thorough evaluation in Section 3.1, `Tai-e`'s pointer analysis system outperforms all other state-of-the-art frameworks in terms of efficiency and soundness, for virtually all cases, under both context-insensitivity and context-sensitivity settings. With such efficient capabilities, other analyses, including those that interact with pointer analysis (Section 4.1) and those that use its results (Section 4.2), can profit from increased analysis speed. In addition, `Tai-e` also offers efficient analysis-relevant data structures (such as index map, various bit sets, and hybrid sets, etc.) to make it easier to develop efficient new analysis.

***Adaptive.*** Build a system that is adaptive to functional changes, and the keys to enabling it are modularity and extension points in design [37]. We create `Tai-e` with adaptivity in mind at all times. For instance, as introduced in Section 3.1, `Tai-e` supports fine-granularity modules for pointer analysis system by providing various kinds of heap abstractions and context-sensitivity approaches, and it offers interfaces to easily implement new pointer analysis algorithms, or change any heap and context components of existing ones as needed. Take another example, to add a new analysis (Section 4.2) in `Tai-e`, users only need to extend a related class (method-/class-/program-level), completes its logics accordingly and register the analysis in a configuration file (no other hard-code operations are required), and `Tai-e` will handle the rest of activities for managing it (e.g., resolving its dependences, driving it with specified options, accessing its results, etc.). Such decouple design improves `Tai-e`'s ability to adjust to the addition or removal of analysis.

***Dependable.*** A system is dependable if it is reliable, available (in case of failures) and secure [37]. However, `Tai-e` is not a software that offers continuous services or stores private user data, and thus dependability is not its major concern at the current stage. Still, we have made some attempts in design to make `Tai-e` more reliable by preventing certain faults that users unwittingly cause, e.g., the IR of `Tai-e` is immutable and the dependencies among multiple analyses (which affects execution order of analyses) are resolved automatically by `Tai-e` based on their configurations (rather than having users specifically list them in input option). In addition, inspired by TAJS (a classic static analysis framework for JavaScript) [26], `Tai-e` supports regression testing which can take advantage of the integrated analyses to assess the correctness of framework code in the event of any improper modifications.

***Yummy.*** A system should be yummy, which implies it ought to have an attractive feature that customers are enthusiastic about [37]. In our opinion, one yummy design of `Tai-e` is its analysis plugin system which facilities the development of analysis that needs to interact with pointer analysis (Section 4.1). To make a yummy system, we'd better study the users deeply, and "it is much easier if the designers are also users" [37]. That exactly describes our case: we have been researching pointer analysis and the analyses rely on it for about ten years, and we know users' needs of pointer analysis well. As described in Section 4.1, a dozen analyses including fundamental ones, clients, utility tools, language features and runtime environment modeling, are all developed on top of this system (actually, it enables many analyses to be pluggable, which introduces fine modularity to Tai-e to make it more adaptable). Moreover, our analysis plugin system is currently being used by a number of ongoing internal projects implemented by different developers (these projects will be released when finished), and the feedback from developers is very promising: everyone agrees that it can fulfill their practical needs and is simple to understand and apply.

---

[2] Note that if a principle or certain aspects of it have no bearing on the primary goal of `Tai-e`, for examples, "Timely" means "shipping soon enough to meet time-to-market needs", and a system should be "Adaptive" to changes "in scale" (from 100 to 1 million users), we will omit the related discussion.

## 7 Related Work

The most pertinent work has been thoroughly compared and discussed throughout the paper, and in this section, we discuss additional related work.

Lam et al. [35] give a retrospective of Soot by summarizing its main features, major changes (e.g., it consolidated singletons to support multiple runs, and supported to analyze incomplete programs), and future directions (e.g., to build faster startup and enhance interprocedural analysis). In addition, they mention some difficulties in developing Soot and suggest some desirable features for future compiler frameworks. Particularly, the authors observe that it is difficult to publish framework papers and urged the community to embrace more framework articles; their appeal is a great encouragement to us.

Schubert et al. [63] introduce the lessons from constructing a data flow analysis framework for C/C++ [62], some of which are specific to framework's features while others are more general. For instance, it would be beneficial for a framework to offer means like instrumentation to help debug analysis-related bugs. This is inspiring and we may consider developing similar approach in Tai-e in the future.

Sadowski et al. [61] summarize the lessons from building static analysis tools at Google. They advise integrating static analysis into workflow as early as possible, and do the analysis checks as compiler errors if possible (otherwise, developers often ignore analysis results). Integrating Tai-e's individual analyses into developers' workflows may need to modify the basic infrastructure of Tai-e, but that is an interesting subject that merit further investigation.

Some studies evaluate static analysis tools from the view of users' needs [15, 16, 55]. For instance, a good static analysis tool should produce high-quality warning messages that offer information on what might be wrong, why it should be fixed, and how it could be fixed, have low false positives of analysis results, and support the integration of user knowledge, and more.

Although the work mentioned above offers helpful lessons (from building specific static analysis frameworks), or study results (by evaluating static analysis tools based on users' requirements), none of them addresses the problem of our research. We still lack a systematic way to observe and discuss a static analysis framework's design quality from the perspective of the developers who build their analyses using the abstractions and facilities provided by the framework. Our work substantially fills this gap by comprehensively compare and discuss the designs of various classic frameworks for Java. Below we briefly describe some other Java static analysis framework.

Chord [56] is a static analysis framework for Java that is written in Java and Datalog with bddbddb [79] (a BDD-based implementation of Datalog), serving as the Datalog solver. As explained in Section 4.1, despite being elegant when implementing various analyses, Datalog has limited expression capacity and optimization potential. Chord is particularly known for its capability to detect concurrent errors such as data races, and we will develop these clients imperatively (rather than declaratively) in Tai-e.

OPAL [17] is a static analysis framework for Java which is written in Scala. The collaborative analysis approach [23] described in Section 4.1 is implemented in OPAL, but unfortunately their approach is too complex to be effective for our problem: developing analysis that interacts with pointer analysis. In addition, the same authors [60] conduct interesting study to assess the soundness of call graphs produced by call-graph algorithms (like RTA and CHA) in various frameworks, emphasizing the importance of effectively handling language features.

We discuss TAJS [26], a classic static analysis framework for JavaScript and Node.js [18, 53, 57], as some designs of Tai-e are inspired by it. One is the regression testing mentioned in Section 6 and the other is the initial idea of Tai-e's analysis plugin system, where the solver-plugins structure resembles TAJS's monitor approach, despite that their goals, methodologies and APIs are fundamentally different. For examples, their monitor approach is mainly used to collect analysis results and perform statistics, e.g., recording the times a statement is accessed, or timing an analysis spent on each statement, and it requires that monitor interface implementations should not have side effects on the analysis state prior to the monitor scan, but Tai-e's interface implementation for analysis plugin is to have side effect on the pointer analysis by calling methods like addPointsToSet as explained in Section 4.1.

## 8 Conclusions

From the perspective of analysis developers, what should a good static analysis framework look like? Even after years, there might not be a satisfactory answer to this subject question. But in the real world, it is a valuable question that cannot be avoided as a static analysis framework serves as the foundation for creating static analyses for various applications. To bridge this knowledge gap, this paper takes a big step forward by systematically comparing and discussing the design trade-offs for each of the crucial components of a static analysis framework for Java including program abstraction, fundamental analyses, new analysis development and multiple analysis management. For each component, the design choice made for it by each of the related classic frameworks (Soot, WALA, Doop and SpotBugs) are deeply examined and debated according to the HBDC principle.

Our efforts are both labor- and intelligence-intensive as for each design point, we must study and comprehend the code of those large and intricate frameworks full of challenging analysis algorithms. But such efforts are also worthwhile as they aid in the creation of Tai-e, a new static analysis framework for Java that features arguably the "best" designs from both the novel ones we proposed and those of classic frameworks. Tai-e was meticulously constructed from the ground up with the intention of being easy-to-learn and easy-to-use, and we have demonstrated throughout the article that it works well in attaining the goals. Although we cannot ensure to make a comprehensive and accurate assessment for every framework design point (due to the subjective nature of this problem), we believe this paper and Tai-e provide useful materials and perspectives for building better static analysis infrastructures, and we will actively and constantly contribute to Tai-e by developing and incorporating more analyses and clients in the future.

[71] Manu Sridharan, Stephen J. Fink, and Rastislav Bodík. 2007. Thin slicing. In *Proceedings of the ACM SIGPLAN 2007 Conference on Programming Language Design and Implementation, San Diego, California, USA, June 10-13, 2007*, Jeanne Ferrante and Kathryn S. McKinley (Eds.). ACM, 112–122. https://doi.org/10.1145/1250734.1250748

[72] Tamás Szabó, Gábor Bergmann, Sebastian Erdweg, and Markus Voelter. 2018. Incrementalizing Lattice-Based Program Analyses in Datalog. *Proc. ACM Program. Lang.* 2, OOPSLA, Article 139 (oct 2018), 29 pages. https://doi.org/10.1145/3276509

[73] Tian Tan, Yue Li, Xiaoxing Ma, Chang Xu, and Yannis Smaragdakis. 2021. Making Pointer Analysis More Precise by Unleashing the Power of Selective Context Sensitivity. *Proc. ACM Program. Lang.* 5, OOPSLA, Article 147 (oct 2021), 27 pages. https://doi.org/10.1145/3485524

[74] Tian Tan, Yue Li, and Jingling Xue. 2016. Making k-Object-Sensitive Pointer Analysis More Precise with Still k-Limiting. In *Static Analysis - 23rd International Symposium, SAS 2016, Edinburgh, UK, September 8-10, 2016, Proceedings (Lecture Notes in Computer Science, Vol. 9837)*, Xavier Rival (Ed.). Springer, 489–510. https://doi.org/10.1007/978-3-662-53413-7_24

[75] Tian Tan, Yue Li, and Jingling Xue. 2017. Efficient and precise points-to analysis: modeling the heap by merging equivalent automata. In *Proceedings of the 38th ACM SIGPLAN Conference on Programming Language Design and Implementation, PLDI 2017, Barcelona, Spain, June 18-23, 2017*, Albert Cohen and Martin T. Vechev (Eds.). ACM, 278–291. https://doi.org/10.1145/3062341.3062360

[76] Frank Tip. 1995. A survey of program slicing techniques. *J. Prog. Lang.* 3, 3 (1995). http://compscinet.dcs.kcl.ac.uk/JP/jp030301.abs.html

[77] Raja Vallée-Rai, Phong Co, Etienne Gagnon, Laurie J. Hendren, Patrick Lam, and Vijay Sundaresan. 1999. Soot - a Java bytecode optimization framework. In *Proceedings of the 1999 conference of the Centre for Advanced Studies on Collaborative Research, November 8-11, 1999, Mississauga, Ontario, Canada*, Stephen A. MacKay and J. Howard Johnson (Eds.). IBM, 13. https://doi.org/10.1145/781995.782008

[78] WALA. 2018. Watson Libraries for Analysis. http://wala.sf.net.

[79] John Whaley and Monica S. Lam. 2004. Cloning-based Context-sensitive Pointer Alias Analysis Using Binary Decision Diagrams. In *Proceedings of the ACM SIGPLAN 2004 Conference on Programming Language Design and Implementation* (Washington DC, USA) *(PLDI '04)*. ACM, New York, NY, USA, 131–144. https://doi.org/10.1145/996841.996859

[80] YAML. [n.d.]. https://yaml.org/.
Tian Tan and Yue Li18